# Time-dependent quantum Monte Carlo: preparation of the ground state

**I P Christov**
Physics Department, Sofia University, 1164-Sofia, Bulgaria
E-mail: ipc@phys.uni-sofia.bg

**Abstract.** We study one-dimensional (1D) and two-dimensional (2D) Helium atoms using a new time-dependent quantum Monte Carlo (TDQMC) method. The TDQMC method employs random walkers, with a separate complex-valued guiding wave attached to each walker. The ground state is calculated by a self-consistent solution of complex-time Schrödinger equations for the guiding waves and of equations for the velocity fields of the walkers. Our results show that the many-body wavefunction and the ground state energy of the model atoms are very close to those predicted by the standard diffusion quantum Monte Carlo method. The obtained ground state can further be used to examine correlated time-dependent processes which include, for example, interaction of atoms and molecules with external electromagnetic fields.

## 1. Introduction

The many-electron problem is one of the most intractable problems of quantum physics. The great difficulty is that the motions of the electrons in atoms and molecules are correlated because of the strong Coulomb repulsion between their negative charges. The solution of the many-electron Schrödinger equation is therefore a complicated function of $3N$ variables, where $N$ is the number of electrons in the system. The introduction of density functional theory (DFT) made it possible to reformulate theory in such a way that the important physical quantity is the electron density, rather than the wavefunction, reducing the dimensionality of the problem to three [1]. In this way in DFT the many-body problem is effectively relocated into the definition of the exchange-correlation functional, whose mathematical expression as function of the electron density is not currently known and unlikely ever to be known exactly. The many-electron problem becomes even more complex when time dependent processes in external fields are involved in the dynamics, e.g. where short-pulse ionization and/or charge transfer take place. Such processes become increasingly important with the advent of attosecond-duration soft x-ray laser sources [2,3] which made it possible to conduct experiments with unprecedented time resolution. Clearly, the time-dependent Hartree–Fock approximation, which neglects electron correlation, fails to reproduce the correct correlated electron dynamics. The time-dependent density functional theory (TDDFT) offers a possibility to include the correlation effects in a numerically tractable way [4]. Parametrized functionals are constructed which allow the calculation of ionization probabilities from the time-dependent density alone. However, it has been shown that the time-dependent exchange-correlation potentials may not be sufficient to describe the correct electron correlated dynamics in time-varying fields. It was pointed out [5] that not only the density at a given time but its whole history enters the exchange-correlation term in a time-dependent calculation. If the history of the density is ignored, ambiguities are introduced into the representation of excited states. More recently, progress has been achieved by multiconfigurational time-dependent Hartree [6] (-Fock [7]) method which allows the inclusion of electron correlation effects in a more systematic way.

A completely different approach to the electron correlation problem is offered by the quantum Monte Carlo (QMC) techniques, a scheme which in principle can yield the exact ground state energy and wavefunction of a complex quantum system [8]. For example, the diffusion QMC is based on the similarity between the imaginary time solution of the Schrödinger equation and a generalized diffusion stochastic motion of Brownian particles (walkers), where one can compute the value of the wavefunction by following the motion of these walkers in configuration space. In order that higher accuracy is attained it is necessary to guide the walkers using a guide function that well approximates the exact wavefunction. Often, variational QMC is used as a fist approximation since the trial energy is close to the exact one and the wavefunction accurately approximates the ground state wavefunction. The QMC represents the most accurate method currently available for medium-sized and large systems. Standard QMC calculations scale as the third power of the number of electrons (the same as DFT), and are capable of treating molecules as well as solid crystalline phases [9].

However, the present day QMC techniques suffer from a serious drawback in that they cannot describe time dependent processes, e.g. in atoms and molecules exposed to an external field, including charge transfer and ionization. One of the underlying reasons is that in QMC the stochastic component in the walker's motion (the creation/destruction of configurations) is dominating the evolution of the system while the guide function plays a secondary role. Recently, different approach was proposed [10] where separate complex-valued guide function is attributed to each individual walker and these guide functions evolve according to the time dependent Schrödinger equation, concurrently with the motion of the corresponding walkers. In this way, using notions from QMC and quantum fluid dynamics with trajectories, the guide waves and the quantum fluid particles (walkers) participate in the dynamics on an equal footing within the new formalism. In this paper we consider the preparation of the atomic ground state for two model atoms within the frames of the new time-dependent quantum Monte Carlo (TDQMC) technique.

## 2. Model and method

For N-electron atom, the non-relativistic time-dependent Schrödinger equation reads:

$$i\hbar\frac{\partial}{\partial t}\Psi(\mathbf{r}_1,...,\mathbf{r}_N,t)=\left[\sum_{i=1}^{N}\left(-\frac{\hbar^2}{2m}\nabla_i^2\right)+V_{e-n}(\mathbf{r}_1,...,\mathbf{r}_N)+V_{e-e}(\mathbf{r}_1,...,\mathbf{r}_N)\right]\Psi(\mathbf{r}_1,...,\mathbf{r}_N,t) \quad (1)$$

Here $V_{e-n}(\mathbf{r}_1,...,\mathbf{r}_N)$ is the electron-nuclear potential and $V_{e-e}(\mathbf{r}_1,...,\mathbf{r}_N)$ is the electron-electron potential. In the standard diffusion quantum Monte Carlo (DMC) we first pick an ensemble of configurations chosen from a certain distribution. This ensemble is next evolved according to the short-time approximation to the Green function of equation (1), for imaginary time. Within this approximation the Green function separates into two processes: random diffusive jumps of the configurations arising from the kinetic term, and creation/destruction of configurations arising from the potential energy term [11]. In presence of a guide function, an additional drift of the particles is introduced that is proportional to the gradient of the guide function, as the guide function is the same for many particles. In TDQMC approach considered here each walker from the ensemble is attached to its own complex-valued guide function which evolves in time, concurrently with the evolution of the trajectory of that walker. Thus, our approach combines the viewpoints of quantum Monte-Carlo and quantum trajectory dynamics within a new framework for doing quantum calculations, where a set of random walkers is identified with fluid particles whose motion is governed by the equations of quantum fluid dynamics [12,13]. However, the equations of the standard quantum fluid dynamics are nonlinear and contain quantum potentials, which can be an obstacle for their robust numerical treatment. Therefore, we adopt a modified approach which reduces the many-body problem to a set of coupled Schrödinger equations for the guide waves, and equations for the velocity field of the walkers. This is accomplished by using a representation of the many-body wavefunction as a polar decomposition of single-particle real-valued amplitudes and a many-body phase term $\Psi(\mathbf{r}_1,...,\mathbf{r}_N,t)\approx R_1(\mathbf{r}_1,t)...R_N(\mathbf{r}_N,t)\exp[iS(\mathbf{r}_1,...,\mathbf{r}_N,t)/\hbar]$. The main physical motivation behind such representation is that in the de Broglie-Bohm mechanics the particle motion is determined by the gradient of the many-body phase function (see [13] and equation (A.6) below), while the factorization of the many-body amplitude that we use serves to simplify the equations for the guiding waves. Thus, the synthesis between QMC and quantum hydrodynamic concepts has led to the assumption that each electron is represented by an ensemble of M walkers, each of those moving along certain quantum trajectory determined by the evolution of the guide function. The guide function $\varphi_i^k(\mathbf{r}_i,t)$ for the *k*-th walker from the *i*-th electron ensemble evolves according to the time dependent Schrödinger equation (see the appendix and [10]):

$$i\hbar\frac{\partial}{\partial t}\varphi_i^k(\mathbf{r}_i,t)=\left[-\frac{\hbar^2}{2m}\nabla_i^2+V_{e-n}(\mathbf{r}_i)+\sum_{\substack{j=1\\ j\neq i}}^{N}V_{e-e}[\mathbf{r}_i-\mathbf{r}_j^k(t)]\right]\varphi_i^k(\mathbf{r}_i,t), \quad (2)$$

where *i*=1,..,N; *k*=1,2,…M. In equation (2) $\mathbf{r}_j^k(t)$ represents the trajectory of the *k*-th walker from the *j*-th electron ensemble. In this way the many-body Schrödinger equation has been reduced to a set of single-particle Schrödinger equations coupled by a time-dependent potential which includes the instantaneous total Coulomb field experienced by each walker. This time-dependent potential is a sum of the Coulomb fields of the walkers which represent the rest of the electrons (the third term in equation (2)). Note that equation (2) can be derived also from the Hartree theory under some physically motivated assumptions, as was done in [10]. On the other hand, the walkers also experience force that is proportional to the gradient of the corresponding guide function. In conventional DMC that force is a result of the importance sampling procedure, which transforms the Schrödinger equation into a Fokker-Planck-type of equation [8]. In TDQMC complex-valued wavefunctions are used, and therefore the walkers are guided by using the de Broglie-Bohm pilot-wave relation [13]:

$$\frac{d\mathbf{r}_i^k(t)}{dt}=\frac{\hbar}{m}\mathrm{Im}\left[\frac{1}{\Psi(\mathbf{r}_1,...,\mathbf{r}_N,t)}\nabla_i\Psi(\mathbf{r}_1,...,\mathbf{r}_N,t)\right]_{\mathbf{r}_j=\mathbf{r}_j^k(t)}, \quad (3)$$

where i,j=1,…,N; k=1,…,M. It is important to point out that equations (2) and (3) comprise a self-contained set of equations of motion for the walkers and the guide waves. Since the Schrödinger equation (1) does not contain spin variables, the electron statistics is determined by the symmetry properties of the many-body wavefunction under exchange of the arguments. However, in contrast to the Hartree-Fock approximation, the equations for the guiding waves (equation (2)) do not contain exchange terms. Therefore, the exchange interaction between parallel-spin electrons should be accounted for by equation (3), where the many-body wavefunction can be represented as a Slater determinant composed from the guiding functions $\varphi_i^k(\mathbf{r}_i,t)$. In this way, the N-dimensional quantum state may acquire time-dependent nodal surfaces to additionally rule the motion of the walkers. Therefore, in TDQMC formalism the two most important sources of electron correlation, namely the electron-electron repulsion and the exchange-induced correlation, are introduced by the two equations (2) and (3). The repulsion-induced correlation is included through the third term in the wave equations (2), while the exchange-correlation is introduced through the trajectory equations (3).

## 3. Results

It is important to stress that the TDQMC formalism does not involve calculation of integrals, which has been a major difficulty for most of the non-Monte Carlo methods. Although in TDQMC we still calculate wavefunctions, the quantities of physical interest are expressed entirely in terms of walker configurations. For example, the energy of the stationary states of the Hamiltonian can be calculated as a sum over the walkers, under the condition that the walker velocity is zero:

$$E = \frac{1}{M}\sum_{k=1}^{M}\left[\sum_{i=1}^{N}\left[-\frac{\hbar^2}{2m}\frac{\nabla_i^2\varphi_i^k(\mathbf{r}_i^k)}{\varphi_i^k(\mathbf{r}_i^k)} + V_{e-n}(\mathbf{r}_i^k)\right] + \sum_{\substack{i,j=1\\ i>j}}^{N} V_{e-e}(\mathbf{r}_i^k - \mathbf{r}_j^k)\right] \quad (4)$$

To determine the ground state of the system, the standard DMC uses imaginary-time propagation, where the wavefunction remains real-valued at all times. In TDQMC complex time is used instead, which ensures a non-zero velocity of the walkers in equation (3). By working with complex-valued wavefunctions in TDQMC we assume that the walker distribution corresponds to the probability density (modulus squared of the guide function), whereas in DMC the walker distribution corresponds to the modulus of the guide function. It is worth mentioning also that in TDQMC formalism no initial guess for the ground state energy is required because the guide waves evolve inevitably to the ground state of the system. In DMC the guide function is constructed to minimize the number of divergences in the local energy caused by the Coulomb interactions. This is done by using appropriate Slater–Jastrow-type many-body wavefunction which includes free parameters to be optimized [14]. In TDQMC the correlation is accounted for naturally through explicit Coulomb potentials (the third term in equation (2)), with no free parameters involved.

First, we compare the walker motion in DMC and in TDQMC. To this end, one-dimensional and two-dimensional Helium atoms are considered where the two electrons with anti-parallel spin evolve from a given initial state to the ground state. In order to specify the role of the electron-electron correlation we use smoothed Coulomb potentials with different parameters for the electron-nuclear interaction:

$$V_{e-n}(\mathbf{r}_i) = -\frac{2e^2}{\sqrt{a + r_i^2}}\ , \quad (5)$$

and for the electron-electron interaction:

$$V_{e-e}[\mathbf{r}_i - \mathbf{r}_j^k(t)] = \frac{e^2}{\sqrt{b + [r_i - r_j^k(t)]^2}} \qquad (6)$$

The calculation begins with randomly positioned walkers with equal initial guide functions. The time evolution of both the walker's position and the guide wave is determined by a self-consistent numerical solution of the set of equations (2) and (3), where each walker experiences a specific quantum force along its trajectory, and we add a linearly-decreasing in time random component to the walker position in order to thermalize the distribution at each time step. This thermalization is needed to avoid possible bias in the walker distribution for the ground state that may arise due to the quantum drift alone. Additional selection of walkers can be used in this method, by destructing some of the walkers. For simple system like Helium, equations (2) can be integrated numerically in fixed in space grids using Crank-Nicholson method which is unconditionally stable and second order accurate in space and time, while second-order Runge-Kutta method is used for the equations for the walkers (3). Using basis set expansion or finite elements would be beneficial for more complex systems. The instantaneous position of each walker $\mathbf{r}_j^k(t)$ in equation (2) is calculated using complex linear interpolation. It is noteworthy mentioning that besides the Eulerian viewpoint considered here, Lagrangian "go with the flow" viewpoint can also be implemented [15].

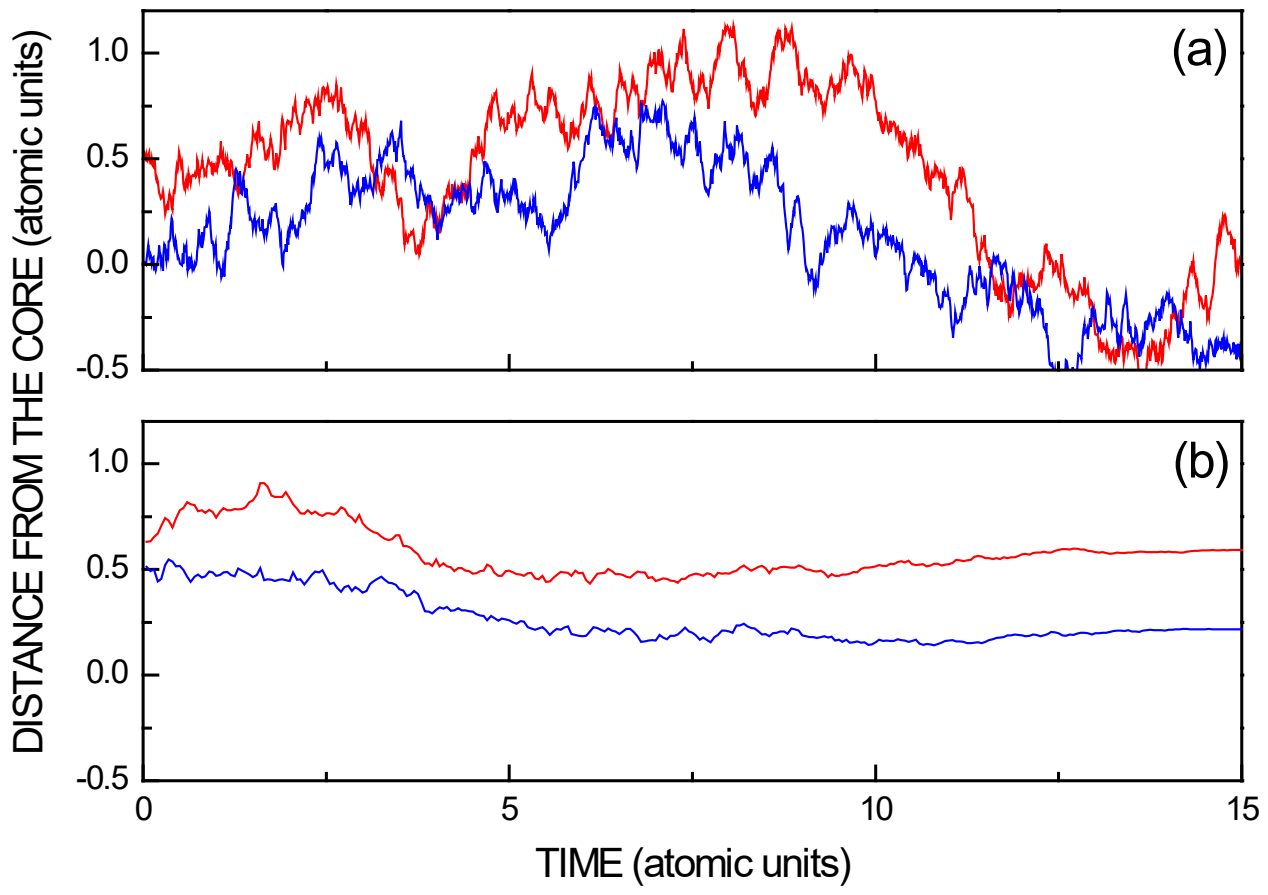

**Figure 1.** Time evolution of the coordinates of two arbitrary walkers for 1D Helium: (a) for DMC and (b) for TDQMC.

First, the ground state of one-dimensional Helium atom is calculated by using DMC and TDQMC techniques. 1D-Helium is very appropriate for comparison of the results from the two models because its configuration space is two-dimensional. We choose the two-body wavefunction to be symmetric with respect to the two electrons, which features equivalent electrons of opposite spin (singlet-spin ground state), i.e. $\Psi(x_1,x_2,t) = \varphi_1(x_1,t)\varphi_2(x_2,t) + \varphi_1(x_2,t)\varphi_2(x_1,t)$. In this simulation 5000 random walkers are used with an initial Gaussian distribution of width 1 atomic unit (a.u) and with all initial guide functions centered at the core. Throughout this paper the smoothing parameters in equations (5) and (6) are $a$=1 a.u. and $b$=0.2 a.u., which corresponds to strongly correlated electrons where electron-electron interaction dominates over electron-nuclear interaction. Spatial grid size of 30 a.u. and complex-time step size (0.05,-0.05) in atomic time-units are used. Figure 1 shows the time evolution of the trajectories of two random walkers from the ensemble, which correspond to the two electrons of the atom for DMC and for TDQMC, after 300 time steps. It is seen from figure 1(b) that the TDQMC trajectories tend to steady-state positions close to the end of the interval, whereas the DMC trajectories (figure 1(a)) experience random jumps at all times. The different behavior in these two cases is due to the essentially different mechanisms for walker guiding used in the two techniques. In TDQMC the guide functions evolve to steady-state where the right-hand side in equation (3) approaches zero, while in DMC the walker coordinates experience random jumps at each time step. The time evolution of the two ensembles is shown in video 1, with final distributions shown in figure 2. In TDQMC calculation the guide functions are normalized to unity at each time step. The predicted ground state energy is -1.942 a.u. for DMC, and -1.936 a.u. for TDQMC. It should be noted that the walker distributions shown in figure 2(a) and 2(b) look very different although they represent two identical ground states. This is because in DMC the walker ensemble approximates the ground state wavefunction, while in TDQMC it corresponds to the probability density.

In order to compare more precisely the two distributions shown in figure 2, we present in figure 3 the contour maps obtained after interpolation of these distributions with Gaussians of width 0.5 a.u., and next squaring up the distribution from figure 2(a) to obtain the probability density. It is seen that the two maps are very similar, and they exhibit the characteristic fluted contours where the dents along the $x_1=x_2$ line evidence the correlated electron motion [16]. We also compare the TDQMC results for the ground state energy with the results from the exact diagonalization of the Hamiltonian for 1D Helium, which yields -1.941 a.u. For the same parameters, the Hartree-Fock approximation gives -1.836 a.u. for the ground state energy. Therefore, we see that a large portion of the ground state correlation energy is taken into account by TDQMC method, whose predictions differ within 1% from the exact value and from the DMC result.

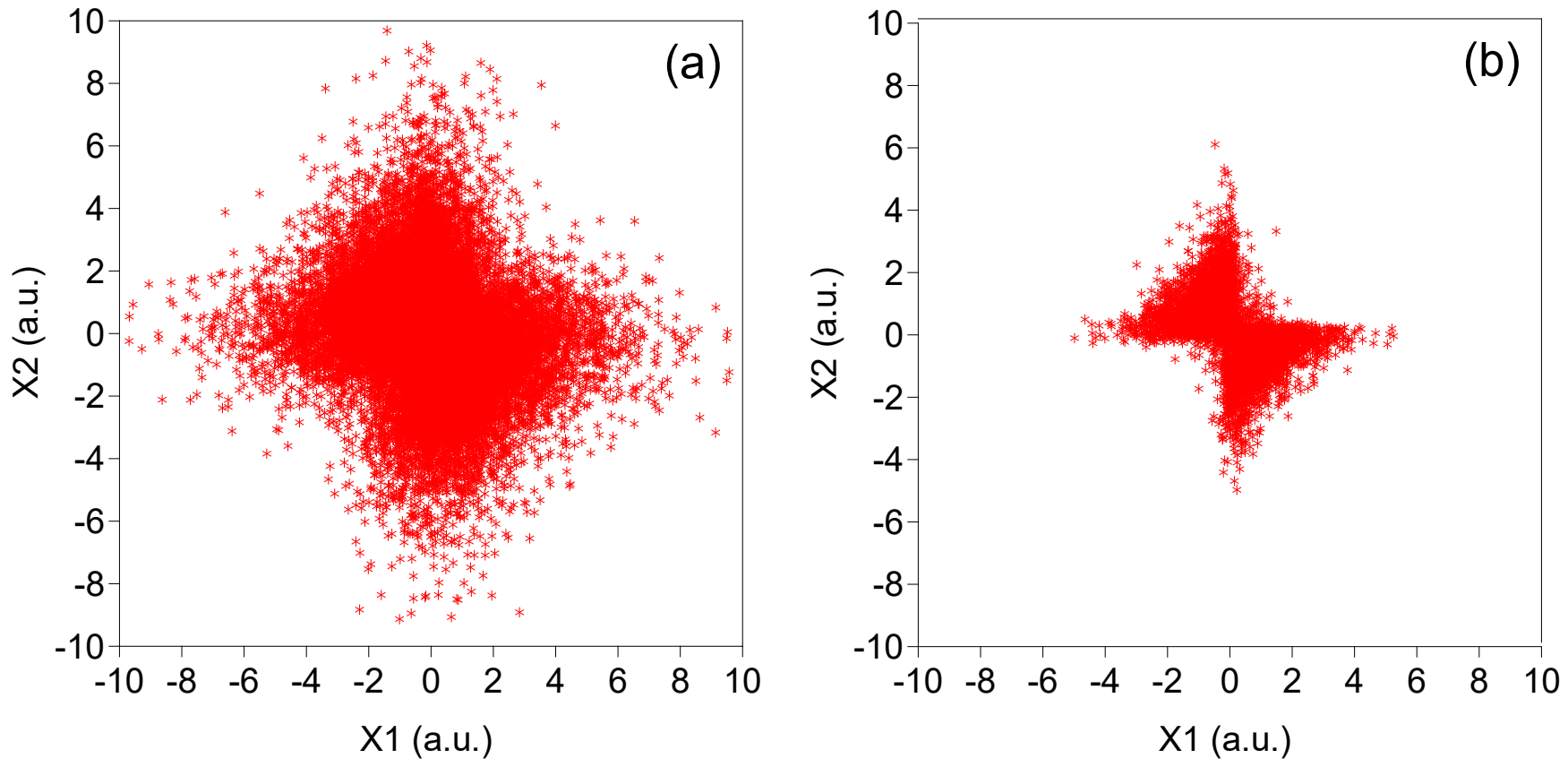

**Figure 2.** Walker distribution in configuration space for 1D Helium: (a) for DMC and (b) for TDQMC. Symmetric ground state is assumed.

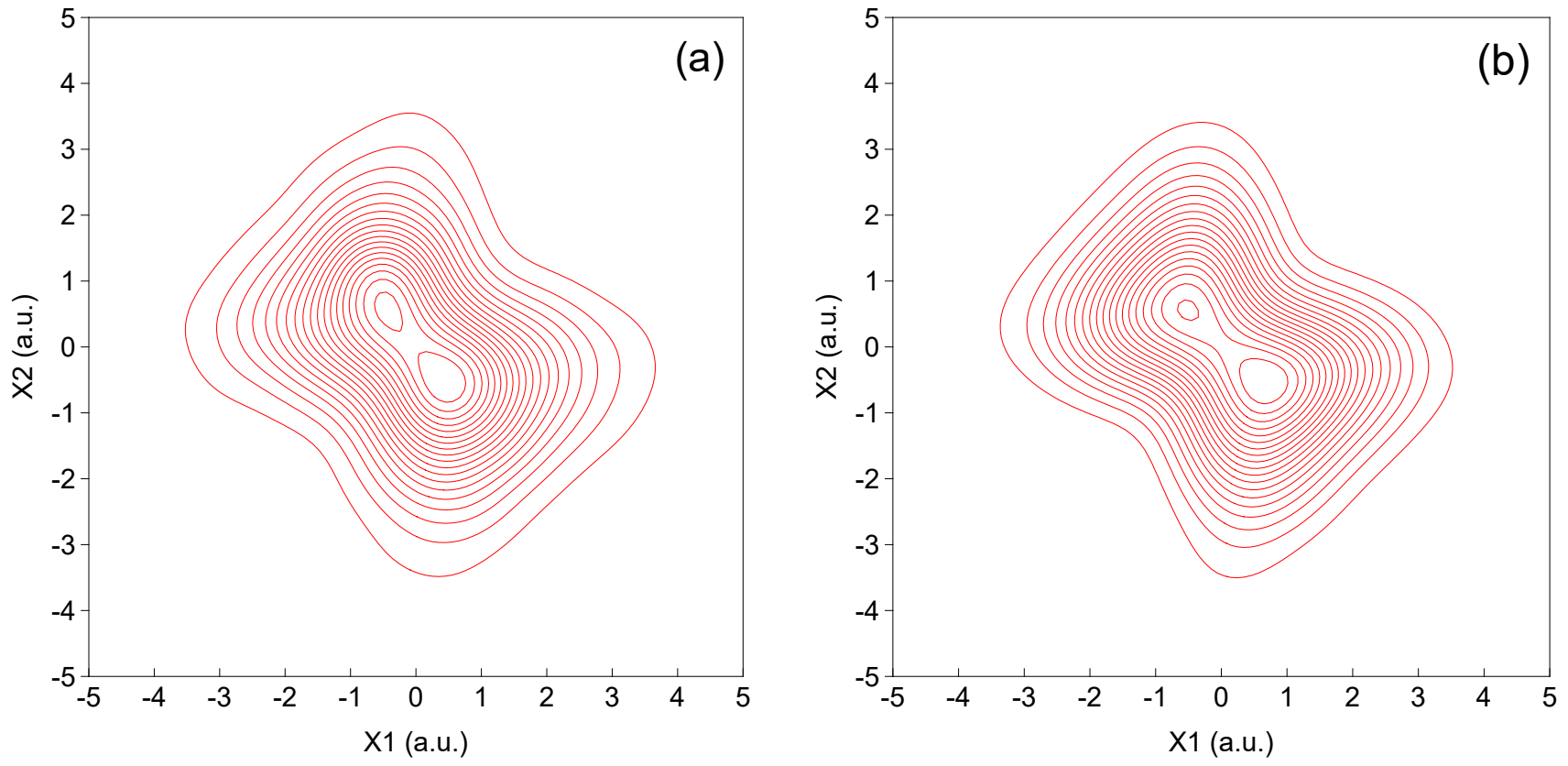

**Figure 3.** Contour maps of the walker distributions from figure 2: (a) for DMC and (b) for TDQMC.

Next, we calculate the walker distribution for the ground state of two-dimensional Helium atom by using TDQMC technique. For $a$=1 a.u. and $b$=0.2 a.u. in equations (5) and (6), the ground state energy predicted by DMC is -1.5155 a.u. while TDQMC gives -1.5012 a.u. Since in this case the configuration space is four-dimensional ($x_1$,$y_1$,$x_2$,$y_2$), we can only show 2D-sections of that distribution. Figure 4 shows four 2D-sections which correspond to the ($x_1$,$y_1$), ($x_1$,$x_2$), ($x_1$,$y_2$), and ($y_1$.$y_2$) walker distributions. It is seen in figure 4(a) that the ($x_1$,$y_1$) distribution is symmetric in all directions, which is expected since the two coordinates of each separate electron are independent. On the other hand, the shapes of the ($x_1$,$x_2$) and the ($y_1$,$y_2$) distributions in figure 4(b) and 4(d) are similar to the 1D case (see figures 2 and 3), but the dents along the $x_1$=$x_2$ and $y_1$=$y_2$ lines are not that seeable. This is due to the additional degree of freedom experienced by the electrons in two dimensions which reduces their correlated interaction to the case where both of their x and y coordinates are close. The square-shape distribution shown if figure 4(c) evidences the degree of correlation between the "crossed" electron coordinates ($x_1$,$y_2$), which is weaker than the correlation between the "uncrossed" coordinates ($x_1$,$x_2$), ($y_1$,$y_2$). This is a result of the specific form of the electron-electron Coulomb potential in Cartesian coordinates.

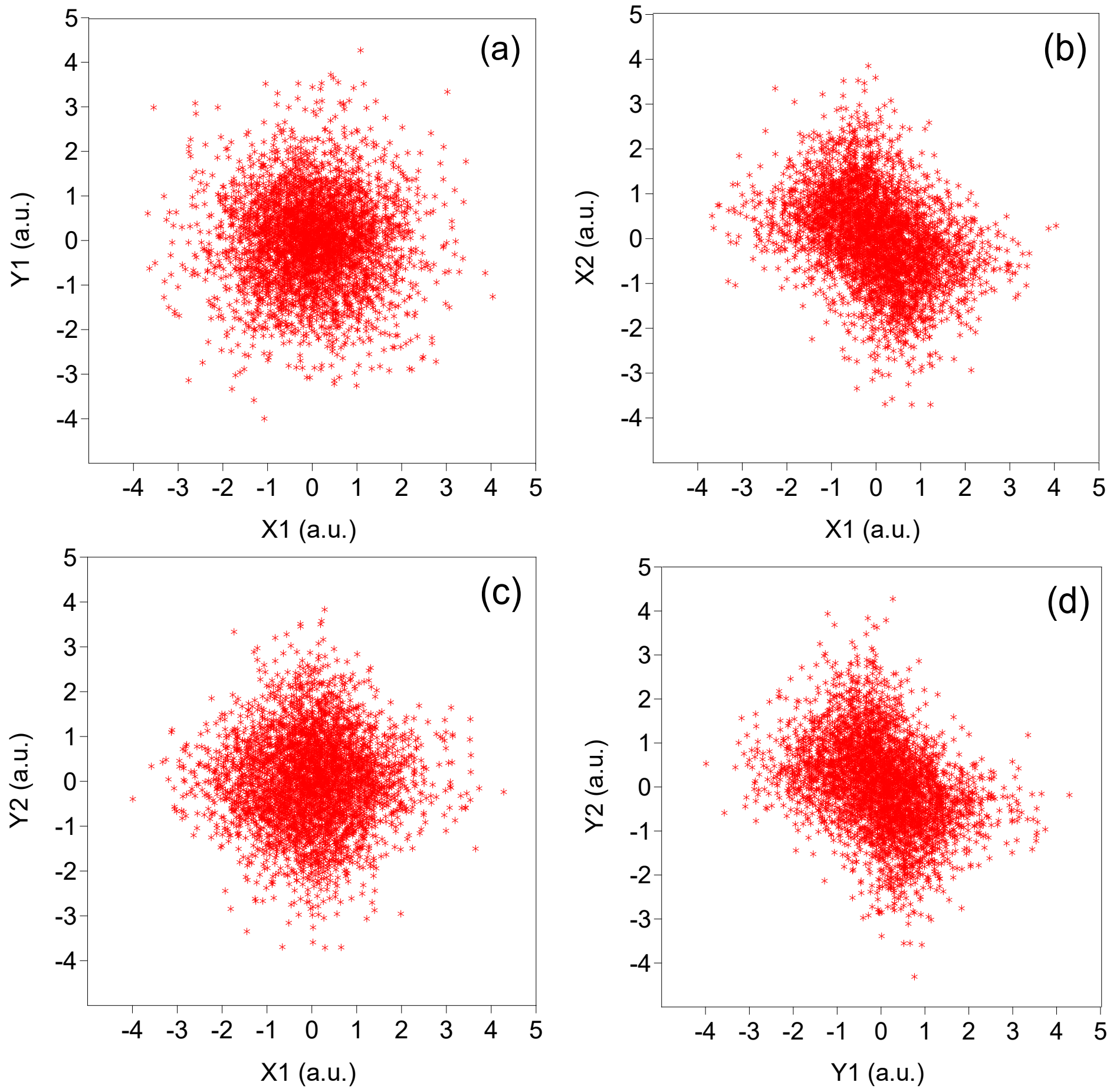

**Figure 4.** Walker distributions in configuration space for 2D Helium (TDQMC result).

For fermionic systems, the ground state wavefunction has nodes where it changes sign. These nodes are essential for the state to be antisymmetric with respect to exchange of any two particles. If no antisymmetry is imposed, the application of the DMC method results in a bosonic solution of lower energy than the true fermionic ground state. The most successful method to address the nodal problem in DMC is the fixed node approximation, followed by an algorithm that releases the nodes of the guide function in order to obtain their correct locations [17]. The key approximation in the fixed-node DMC method is the use of an approximate nodal surface from the guiding function. In practice, the nodes of HF or DFT wavefunctions (Slater determinants) are used. In importance sampling DMC, if the random walk approaches a node of the trial function the walker experiences a "force" which sweeps it away from the node. Similar happens in TDQMC where the nodes occur through the dependence of the many-body wavefunction on the individual time-dependent guide functions (equation (3)). Whenever a walker approaches a nodal surface, the drift velocity in equation (3) grows and carries it away. The latter follows also from the general properties of quantum hydrodynamic trajectories, where the quantum potential keeps these away from the nodal regions and thus they cannot pass through nodes [12,13]. In other words, the antisymmetry of the wavefunction creates a nodal surface that keeps parallel-spin electrons apart from each other.

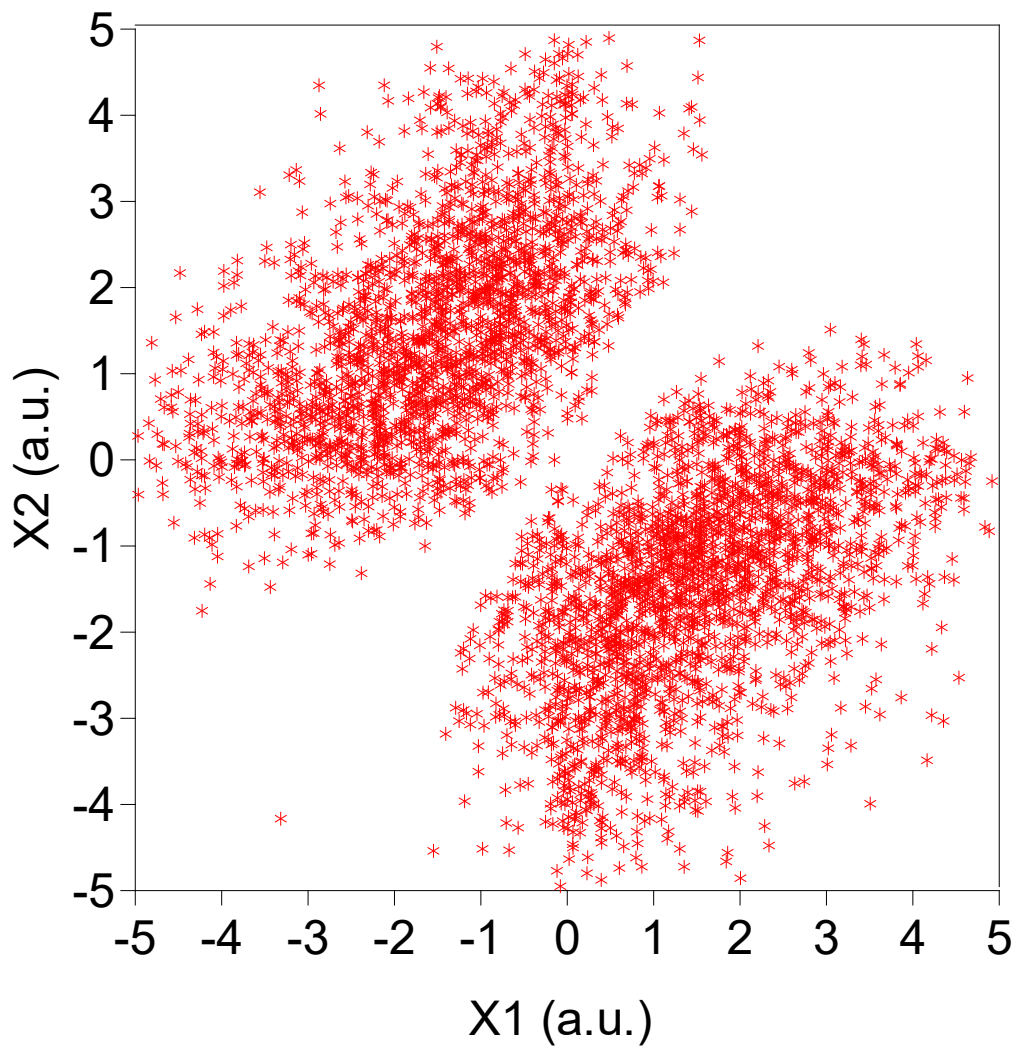

**Figure 5.** Walker distributions in configuration space for 1D Helium with parallel-spin electrons (TDQMC result).

The result from TDQMC calculation for the ground state of 1D Helium with parallel-spin electrons is shown in figure 5, while the time evolution is shown in video 2. The two-body wavefunction in this case is antisymmetric with respect to exchange of the two electrons, $\Psi(x_1,x_2,t)=\varphi_1(x_1,t)\varphi_2(x_2,t)-\varphi_1(x_2,t)\varphi_2(x_1,t)$. It is seen from video 2 that the nodal surface (line in this case) which occurs in $\Psi(x_1,x_2,t)$ pushes the particles away from the region ($x_1$=$x_2$), until the lowest-energy antisymmetric state is established. This calculation can be further improved if the initial walker distribution is prepared by using Metropolis algorithm with an appropriate weight function which is close to the expected probability distribution. The energy of the lowest antisymmetric state we obtained by using TDQMC is -1.761 a.u., which is close to the DMC result of -1.766 a.u. Note that since in TDQMC the population density of walkers is proportional to the modulus square of the complex guiding wave (which is always positive) no "fermion sign problem" occurs in this technique, unlike in conventional DMC.

## 4. Conclusions

In conclusion, we explore a new time dependent quantum Monte Carlo approach where quantum dynamics is modeled by using ensembles of both particles and guiding waves. In this technique, the electron-electron interaction (correlation) is accounted for by using explicit Coulomb potentials, instead of exchange-correlation potentials as is done in the time dependent density functional approximation. This approach can be useful for simulations of very fast (attosecond-duration) strongly correlated phenomena in complex systems such as molecules, clusters, and nanostructures, where fast ionization and/or charge transfer may take place. Other promising applications of TDQMC include the calculation of electromagnetic fields in arbitrary proximity of the quantum objects, well beyond the dipole approximation [18]. Since the calculation of guiding functions is included at each time step, the TDQMC technique is more computationally expensive than the conventional DMC. However, similarly to DMC, the TDQMC algorithm is intrinsically parallel and thus it is easily adapted to parallel computers, with a speedup that scales linearly with the number of processors.

**Acknowledgment**

The author gratefully acknowledges support from the National Science Fund of Bulgaria under contract WUF-02-05.

## Appendix. Derivation of the key equations

We shall now provide a detailed derivation of equations (2) and (3) of the text. We use the notations from Holland's book [13], and we start with the many-body Schrödinger equation (1):

$$i\hbar\frac{\partial}{\partial t}\Psi(\mathbf{r}_1,...,\mathbf{r}_N,t)=\left[\sum_{i=1}^{N}\left(-\frac{\hbar^2}{2m}\nabla_i^2\right)+V_{e-n}(\mathbf{r}_1,...,\mathbf{r}_N)+V_{e-e}(\mathbf{r}_1,...,\mathbf{r}_N)\right]\Psi(\mathbf{r}_1,...,\mathbf{r}_N,t) \tag{A.1}$$

First, we represent the many-body wavefunction as a polar decomposition of product of single-particle real-valued amplitudes, and a many-body phase term, where the phase is assumed to govern the correlated electron motion:

$$\Psi\left(\mathbf{r}_1,...,\mathbf{r}_N,t\right)\approx R_1(\mathbf{r}_1,t)...R_N(\mathbf{r}_N,t)\exp[iS(\mathbf{r}_1,...,\mathbf{r}_N,t)/\hbar] \tag{A.2}$$

We substitute this expression into equation (A.1), which yields:

$$i\hbar\left[\frac{\partial R_1(\mathbf{r}_1,t)}{\partial t}R_2(\mathbf{r}_2,t)...R_N(\mathbf{r}_N,t)+...+R_1(\mathbf{r}_1,t)R_2(\mathbf{r}_2,t)...\frac{\partial R_N(\mathbf{r}_N,t)}{\partial t}\right.$$

$$\left.+\frac{i}{\hbar}R_1(\mathbf{r}_1,t)R_2(\mathbf{r}_2,t)...R_N(\mathbf{r}_N,t)\frac{\partial S(\mathbf{r}_1,...,\mathbf{r}_N,t)}{\partial t}\right]$$

$$=-\frac{\hbar^2}{2m}\left\{\left[\nabla_1^2R_1(\mathbf{r}_1,t)-\frac{1}{\hbar^2}R_1(\mathbf{r}_1,t)\left[\nabla_1S(\mathbf{r}_1,...,\mathbf{r}_N,t)\right]^2+\frac{2i}{\hbar}\nabla_1R_1(\mathbf{r}_1,t)\bullet\nabla_1S(\mathbf{r}_1,...,\mathbf{r}_N,t)+\frac{i}{\hbar}R_1(\mathbf{r}_1,t)\nabla_1^2S(\mathbf{r}_1,...,\mathbf{r}_N,t)\right]\right.$$

$$\times R_2(\mathbf{r}_2,t)...R_N(\mathbf{r}_N,t)+...+R_1(\mathbf{r}_1,t)R_2(\mathbf{r}_2,t)...R_{N-1}(\mathbf{r}_{N-1},t)$$

$$\left.\times\left[\nabla_N^2R_N(\mathbf{r}_N,t)-\frac{1}{\hbar^2}R_N(\mathbf{r}_N,t)\left[\nabla_NS(\mathbf{r}_1,...,\mathbf{r}_N,t)\right]^2+\frac{2i}{\hbar}\nabla_NR_N(\mathbf{r}_N,t)\bullet\nabla_NS(\mathbf{r}_1,...,\mathbf{r}_N,t)+\frac{i}{\hbar}R_N(\mathbf{r}_N,t)\nabla_N^2S(\mathbf{r}_1,...,\mathbf{r}_N,t)\right]\right\}$$

$$+\left[V_{e-n}(\mathbf{r}_1,...,\mathbf{r}_N)+V_{e-e}(\mathbf{r}_1,...,\mathbf{r}_N)\right]R_1(\mathbf{r}_1,t)R_2(\mathbf{r}_2,t)...R_N(\mathbf{r}_N,t), \tag{A.3}$$

where we have canceled the exponential phase term. Next, we take the real part of the two sides of equation (A.3), and after division by $R_1(\mathbf{r}_1,t)...R_N(\mathbf{r}_N,t)$, we arrive at the generalized Hamilton-Jacobi set of equations:

$$\frac{\partial S(\mathbf{r}_1,...,\mathbf{r}_N,t)}{\partial t}+\frac{1}{2m}\sum_{i=1}^{N}\left[\nabla_i S(\mathbf{r}_1,...,\mathbf{r}_N,t)\right]^2-\frac{\hbar^2}{2m}\sum_{i=1}^{N}\frac{\nabla_i^2 R_i(\mathbf{r}_i,t)}{R_i(\mathbf{r}_i,t)}$$

$$+V_{e-n}(\mathbf{r}_1,...,\mathbf{r}_N)+V_{e-e}(\mathbf{r}_1,...,\mathbf{r}_N)=0 \tag{A.4}$$

It is easy to check that the imaginary part of equation (A.3) leads to a continuity equation for the quantum fluid particles (walkers). Next, we apply the operator $\nabla_j \equiv \partial/\partial\mathbf{r}_j$ to both sides of equation (A.4), and after rearranging we get:

$$\left\{\frac{\partial}{\partial t}+\frac{1}{m}\left[\sum_{i=1}^{N}\nabla_i S(\mathbf{r}_1,...,\mathbf{r}_N,t)\bullet\nabla_i\right]\right\}\nabla_j S(\mathbf{r}_1,...,\mathbf{r}_N,t)$$

$$-\frac{\hbar^2}{2m}\nabla_j\frac{\nabla_j^2 R_j(\mathbf{r}_j,t)}{R_j(\mathbf{r}_j,t)}+\nabla_j\left[V_{e-n}(\mathbf{r}_1,...,\mathbf{r}_N)+V_{e-e}(\mathbf{r}_1,...,\mathbf{r}_N)\right]=0 \tag{A.5}$$

Now, we set the velocity fields by using the de Broglie-Bohm relation:

$$\frac{d\mathbf{r}_i}{dt}=\frac{1}{m}\nabla_i S(\mathbf{r}_1,...,\mathbf{r}_N,t), \qquad \text{i=1,…,N} \tag{A.6}$$

and, after isolating a full time derivative along the trajectories in LHS of equation (A.5), we obtain:

$$m\frac{d^2\mathbf{r}_i}{dt^2}=\left\{-\nabla_i\left[-\frac{\hbar^2}{2m}\frac{\nabla_i^2 R_i(\mathbf{r}_i,t)}{R_i(\mathbf{r}_i,t)}+V_{e-n}(\mathbf{r}_i)+\sum_{\substack{j=1\\ j\neq i}}^{N}V_{e-e}(\mathbf{r}_i-\mathbf{r}_j)\right]\right\}_{\substack{\mathbf{r}_i=\mathbf{r}_i(t)\\ \mathbf{r}_j=\mathbf{r}_j(t)}}, \qquad \text{i,j=1,…,N} \tag{A.7}$$

where we have used the result from the action of $\nabla_i \equiv \partial/\partial\mathbf{r}_i$ on the many-body potentials:

$$V_{e-n}(\mathbf{r}_1,...,\mathbf{r}_N)+V_{e-e}(\mathbf{r}_1,...,\mathbf{r}_N)=\sum_{k=1}^{N}V_{e-n}(\mathbf{r}_k)+\sum_{\substack{k,l=1\\ k>l}}^{N}V_{e-e}(\mathbf{r}_k-\mathbf{r}_l) \tag{A.8}$$

The essential difference between equation (A.7) and equation (7.1.8) of [12] is that, by using the factorization (A.2), here we have reduced the sums over the kinetic and potential terms in equations (A.4),(A.8) to a kinetic and potential term for each individual particle. This allows us to look for a single-particle Schrödinger equation which corresponds to the trajectory given by equation (A.7), where $\mathbf{r}_i$ should become an argument of the wavefunction. The new single-particle Schrödinger equation reads:

$$i\hbar\frac{\partial}{\partial t}\varphi_i(\mathbf{r}_i,t)=\left[-\frac{\hbar^2}{2m}\nabla_i^2+V_{e-n}(\mathbf{r}_i)+\sum_{\substack{j=1\\j\neq i}}^{N}V_{e-e}[\mathbf{r}_i-\mathbf{r}_j(t)]\right]\varphi_i(\mathbf{r}_i,t),\qquad \text{i=1,...,N} \tag{A.9}$$

where $\mathbf{r}_j$ (t) is now a parameter which describes the trajectories of the electrons other than the i-th. It can be easily verified that equation (A.7) follows from equation (A.9) using standard polar decomposition for $\varphi_i\left(\mathbf{r}_i,t\right)$ in (A.9), and repeating some of the above steps.

Equation (A.9) coincides with equation (2) of the text, provided that we assign a separate wavefunction (guide function) to each walker (k). On the other hand, equation (A.6) is easily transformed to equation (3).